\documentclass{llncs}
\usepackage{graphicx}
\usepackage{algorithmic}
\usepackage[linesnumbered]{algorithm2e}
\begin{document}
\frontmatter          
\pagestyle{headings}  
\mainmatter              
\title{A Tool for Programming Embarrassingly Task Parallel Applications on CoW and NoW}
%
%
\author{Patrizio Dazzi}
\institute{ISTI - CNR, Italy,\\
\email{patrizio.dazzi@isti.cnr.it},\\
\texttt{http://hpc.isti.cnr.it/\homedir dazzi}}

\maketitle              

\begin{abstract}
Embarrassingly parallel problems can be split in parts that are characterized by a really low (or sometime absent) exchange of information during their computation  in parallel. As a consequence they can be effectively computed in parallel exploiting commodity hardware, hence without particularly sophisticated interconnection networks. Basically, this means Clusters, Networks of Workstations and Desktops as well as Computational Clouds. Despite the simplicity of this computational model, it can be exploited to compute a quite large range of problems. This paper describes JJPF, a tool for developing task parallel applications based on Java and Jini that showed to be an effective and efficient solution in environment like Clusters and Networks of Workstations and Desktops.

\end{abstract}
\section{Introduction}
\label{sec:intro} 

Parallel computing, in a nutshell, is a form of computation in which many computations are performed simultaneously, it is based on the principle that large problems can often be approached by dividing them into smaller ones, which are then solved concurrently.

In parallel computing a problem that can be computed, without a particular effort, separating it into a number of tasks to compute in parallel is generally referred as an embarrassingly parallel problem. Often these tasks have no dependency one each others, hence they tend to require little or no communication of results between tasks, and are thus different from more complex computing problems that may require information exchange between tasks, e.g. the communication of intermediate results. 

Parallel applications realized according to this model are not only easier to implement than more complex kinds of parallel applications; they also do not require high-speed (and expensive) communication infrastructures to scale efficiently when the number of resources involved in the computation increases significatively. As a consequence, typically, this kind of applications are run on clusters, networks of workstations and, more recently, on Clouds and Federation of Clouds. 
Basically these environments consist of infrastructures that allow to deal with a huge amount of computational resources but usually characterized by a limited range of guarantees on network subsystems~\cite{carlini2012cloud,coppola2012contrail}. 
Anyway, in spite of the quite simple structure of this parallel paradigm, several kinds of application can be effectively and efficiently realized according to it. 

Examples include:
\begin{itemize}
\item Distributed relational database queries using distributed set processing
\item Webservers
\item Several fractal calculations, basically all the ones where each point can be calculated independently
\item Brute-force searches in cryptography
\item Large scale image recognition softwares 
\item Computer simulations comparing many independent scenarios, such as climate models
\item Genetic algorithms as well as other evolutionary computation meta-heuristics
\item Numerical weather prediction
\item Simulations of particle physics
\end{itemize}

Not having any particular requirement in terms of data exchange, embarrassingly parallel problem can be computed on server farms built with commodity hardware, which do not have any of the special communication and data storage infrastructure, like the ones used in supercomputers. They are, thus, well suited to large, internet based distributed software platforms, e.g. Condor, BOINC, etc.  

In a previous paper we presented JJPF~\cite{danelutto2005java}, which main features are reported in this paper in Section~\ref{sec:JJPF}. Basically, it consists in a tool for implementing embarrassingly parallel applications, written in Java and exploiting Jini~\cite{ApacheRiver} (formerly known as Jini) for resource discovery and job assignment. 
JJPF provides some interesting features, like:
\begin{itemize}
\item \textit{load balancing} across the computing elements participating in the computation 
\item \textit{automatic resource discovering and recruiting} exploiting standard Jini mechanisms
\item \textit{fault tolerance} achieved by substituting faulty resources with other ones (if any) in a seamless and automatic way. 
\end{itemize}

\section{JJPF}
\label{sec:JJPF}

JJPF provides to programmers a user-friendly tool for programming task parallel application in Java that can be run on Networks or Clusters of Workstations. JJPF basically resemble a master-slaves structure. It is based on a set of distributed slaves providing a stream parallel application computation service. Programmers must write their applications as an arbitrary composition of task farm and pipeline computation patterns. Task farm only applications are directly executed by the distributed slaves, whereas applications made of a composition of task farm and pipeline patterns are automatically pre-processed to get their normal form \cite{pdcs99} and are then submitted to the distributed slaves for their execution.

Using JJPF, programmers can express a parallel computation exploiting the task farm pattern simply using the following two statements:

\vspace{1em}
{{\footnotesize \texttt{BasicClient cm =  new BasicClient(program,null,input,output);}}}

{{\footnotesize \texttt{cm.compute();}}}
\vspace{1em}

Where \texttt{input} (\texttt{output}) is a \texttt{Collection} of input (output) tasks
and \texttt{program} is an array hosting the code that slaves have to compute on their sides. 
The code consists in a \texttt{Class} object relative to the user worker code. Such code must implement a \texttt{ProcessIf} interface. The interface requires the three methods: one to provide the input task data (\texttt{void setData(Object task)}), another one to retrieve the result data (\texttt{Object getData()}) and, finally, a last method to compute results out of task data (\texttt{void run()}). 

This single pair of lines of code indeed defines the parallel computation to be executed, starts its execution and terminates when the parallel execution is terminated. 
JJPF basic architecture uses two kinds of components: clients (consisting in the user programs) and services, namely the instances of the distributed servers that actually compute results out of input task data to execute client programs.

The Algorithms~\ref{alg:client} and~\ref{alg:server} report the pseudo-code of client and service components, respectively. 

The client component recruits available services and forks a control thread for each one of them. The control thread, in turn, fetches the task items to compute from the task vector, delivers them to the remote service and retrieves the computed results, storing them to the result vector. Service recruiting is performed exploiting the support of Jini. The first step consists in finding a lookup service, using standard Jini API, then such service is queried for available services (i.e. the slaves). Each service descriptor obtained from lookup is passed to a distinct control thread. 


\begin{algorithm}
\SetAlgoLined
 \KwResult{Compute an Application exploiting Services}
 
 \textbf{network discovery} of the \textit{LookupService}\;
 \textbf{query} lookup for registered services\;
 \If{services are available}{
 	\ForEach{service}{
  		fork a specific control thread\;
   	}
	\textbf{wait} the end of computation\;
  }
  \textbf{terminate} the program\;
 
\caption{Client side of JJPF} \label{alg:client}
\end{algorithm}

\begin{algorithm}

\caption{Server side of JJPF} \label{alg:server}

\SetAlgoLined
 \KwResult{Compute an Application exploiting Services}
 
 \textbf{network discovery} of the \textit{LookupService}\;
 \While{\textbf{not} terminated}{
	\textbf{register} into lookup\;
 	\textbf{wait} for requests\;
	\textbf{unregister} from the lookup\;
  }
  \textbf{terminate} the program\;
 
\end{algorithm}

The behavior of the services is reported in Algorithm~\ref{alg:server}. Basically, each service registers its own descriptor to the Jini lookup and waits for incoming client requests. Once a request is received, it assumes to be recruited by the client that issued it. Then the service un-registers itself from the lookup and starts serving the task computation requests of the client. 
It is easy to see that this implies that each service serves only a single client.

In order to use JJPF on a workstation network or cluster, just the following three steps have to be performed:
\begin{enumerate}
\item Jini has to be installed and configured,
\item JJPF services has to be started at the machines that will eventually be used to run the JJPF distributed server 
\item  a JJPF client such as the one sketched above has to be prepared, compiled and run on the user workstation.
\end{enumerate}

The key concept in JJPF is that resource discovery is automatically performed in the client run time support. No code dealing with service discovery or recruiting is to be provided by application programmers.

This happens because JJPF strongly relies on the Jini technology and inherits its features. The Jini technology is indeed suitable for running on workstation clusters within local area networks. 

JJPF uses two distinct mechanisms to recruit services to clients. One synchronous and one asynchronous (in fact it consists in a sort of publish-subscribe approach). The synchronous mechanism directly queries the Lookup Service about the Service Ids of the available services, i.e. of the slaves currently running the JJPF. The asynchronous mechanisms works by registering to the Lookup Service an \textit{observer} object that will  alert the client of in case new services becoming available, so that they can be recruited. 

JJPF achieves automatic load balancing among the recruited services, due to the scheduling approach adopted in the control threads managing the remote services. Each control thread fetches tasks to be delivered to the remote nodes from a centralized, synchronized task repository. JJPF also automatically handles faults in service nodes. That is, it takes care of the tasks assigned to a service node in such a way that in case the node does not respond any more they can be rescheduled to other service nodes.

This is possible because, as we introduced before, the only kind of parallel applications that are supported in JJPF, are the ones relying on stream parallel computations. In this case, there are natural \textit{descheduling points} that can be chosen to restart the computation of one of the input tasks, in case of failure of a service node. A trivial one is the start of the computation of the task. Provided that a copy of the task data is kept on the client side, the task can be rescheduled as soon as the control thread understands that the corresponding service node has been disconnected or it is non responding. This is the choice we actually implemented in JJPF, inheriting the design from \texttt{muskel} \cite{danelutto2006joint2,danelutto2006joint,aldinucci2001muskel}. 

\section{Related work}
\label{sec:related}

%

Another of our previous developed structured, parallel programming environment \texttt{muskel} already provides automatic discovery of computational resource in the context of a distributed workstation network. \texttt{muskel} was based on plain RMI Java technology, however and the discovery was simply implemented using multicast datagrams and proper discovery threads. 
The \texttt{muskel} environment also introduces the concept of \textit{application manager} that binds computational resource discovery with autonomic application control in such a way that optimal resource allocation can be dynamically maintained upon specification by the user of a performance contract to be satisfied \cite{danelutto2006joint2,danelutto2006joint,aldinucci2001muskel}.
Several other researchers proposed or currently propose environments supporting stream parallel computations on workstation networks and clusters.  Among the others, we mention Cole's \textit{eskel} library running on top of MPI \cite{eskel}, Kuchen's C++/MPI  skeleton library \cite{kuchen-euro} and $CO_2P_2S$ from the University of Alberta \cite{coopps}. 
The former two environments are libraries designed according to the algorithmic skeleton concept. 
The latter is based on parallel design patterns. 
Several papers are related to PageRank Algorithm, Haveliwala \cite{haveliwala99efficient} explores memory-efficient computation, in \cite{kamvar03extrapolation}. Kamvar et al. discuss some methods for accelerating PageRank calculation and in \cite{gleich04linear} Gleich, Zhukov and Berkhin demonstrate that linear system iterations converge faster than the simple power method and are less sensitive to the changes in teleportation.  
Rungsawang and Manaskasemsak in \cite{DBLP:conf/pvm/RungsawangM03} e \cite{DBLP:conf/aina/ManaskasemsakR04} evaluate the performance supplied by an approximated PageRank computation on a Cluster of Workstation using a low-level peer-to-peer MPI implementation.

%
%
%

\section{Conclusions and Future Work}
\label{sec:conclu}
We described JJPF, a framework supporting the execution of stream parallel application on cluster or networks of workstations. The framework exploits plain Java technology, using Jini to address resource discovery and task assignment. 

Resources are discovered and recruited automatically to compute user applications. Fault tolerance features have been included in the framework such that the execution of a parallel program can transparently resist to node or network faults. Load balancing is guaranteed across the recruited computational resources, even in case of resources with fairly different computing capabilities.

JJPF can be used as a building block for more complex parallel environments, like it happened for PAL~\cite{danelutto2008pal,pal-IW06,pal-arxiv}. In the future we plan to adopt it again, possibly in different kind of scenarios and environment. 

We are currently working to a new version of JJPF that will include a support for optimizing task execution on multicore processors as well as the introduction of futures for reducing the number of thread required on client side to manage the computation.

\bibliographystyle{plain}
\bibliography{bibliodazzi}
\end{document}